# Molecular Configurations Around Single Vacancy in Solid $CO_2$ at 0, 100, and 200 K Studied by Monte Carlo Simulation

Koji Kobashi

Former Research Assistant at the Physics Department, Colorado State University, Fort Collins, CO, USA,

Former Senior Researcher, Kobe Steel, Ltd., Japan, and

Former Special Researcher in Shinko Research Co. Ltd., Japan

31-440, Takasu-cho 2-1, Nishinomiya 663-8141, Japan

Abstract

   The configurations of the molecules nearest to a single vacancy in solid $CO_2$ were studied by the Metropolis Monte Carlo (MC) simulation at temperatures $T$ = 0, 100, and 200 K. It was found that distorted orientational configurations at $T$ = 0 and 100 K became uniform at 200 K. The MC simulation was useful to study the local configurations around the vacancy. This work is a continuation of a precedent paper, arXiv:1711.04976 [cond-mat.mtrl-sci] (2017).

   The present work is a continuation of a precedent article[1] in which molecular configurations in solid $CO_2$ with the *Pa3* structure with and without a single vacancy at zero temperature were investigated using the Monte Caro (MC) simulation technique[2] and a Kihara core potential model for molecular interactions.[3] A cubic basic cell without a vacancy contained 2048 molecules, to which the periodic boundary condition was applied. It was found for the $CO_2$ crystal with a single vacancy in the cubic basic cell that the orientational changes were greater for the three nearest-neighbor (NN) molecules and the equivalent three other molecules located in close proximity to the positions of the oxygen atoms associated with a $CO_2$ molecule that had been removed to make a vacancy at the origin. In the present work, the Metropolis MC simulations were undertaken for the $CO_2$ crystal with a single vacancy at temperatures of 0, 100, and 200 K. The major interest of the present research is to see the influence of temperature on the configurational changes of the 12 molecules nearest to the vacancy. For a reference, similar simulations were also done for the $CO_2$ crystal without vacancy.

   The simulation procedure in the present work was partly different from that in the precedent article.[1] In Ref. 1, the initial crystal structure was *Pa3*, and both the center of mass (CM) positions and the orientations of the molecules were initially randomized within a CM position range of -0.2 Å < $\Delta x$, $\Delta y$, $\Delta z$ < 0.2 Å (1 Å = 0.1 nm) and an orientational range of -20° < $\Delta\theta$, $\Delta\phi < 20°$. These are collectively expressed as $\{\Delta r, \Delta\psi\} = \{0.2 \text{ Å}, 20°\}$, hereafter. The values of $\{\Delta r, \Delta\psi\}$ were decreased stepwise to $\{0.001 \text{ Å}, 0.1°\}$ as the computations were repeated, and the last 20 data were chosen for analysis after the total energy had been stable. A single computational job contained 10,000 rounds of calculations over all molecules, and the acceptance ratio was 0.1 %. In the present work, the MC simulations were undertaken at $T = 0$, 100, and 200 K. The computational procedure was as follows: the initial crystal structure was $Pa3$; the CM positions and the orientations of the molecules were randomized within $\{\Delta r, \Delta\psi\} = \{0.2 \text{ Å}, 30°\}$; the randomized structure returned approximately to the $Pa3$ structure in a single computational job; in passing through computational jobs with $\{\Delta r, \Delta\psi\} = \{0.1 \text{ Å}, 15°\}$, the simulations were repeated with $\{\Delta r, \Delta\psi\} = \{0.1 \text{ Å}, 10°\}$ until the total energy was stable; and the simulations were repeated for more than 60 times to use last 40 data for analysis. In the final stage, $\{\Delta r, \Delta\psi\}$ were set to be larger than those in Ref. 1 so that larger molecular motions were allowed at finite temperatures. A single computational job contained 100,000 rounds of calculations over all molecules, and the acceptance ratio was only about 0.005 %. The major reason for the extremely low acceptance ratio was due to the large values for $\{\Delta r, \Delta\psi\}$, and hence new molecular configurations were mostly rejected. Each job took about 70 minutes. The total energy of the $CO_2$ crystal reached the equilibrium after about 20 jobs from the start.

The potential model used are described in Refs. 1 and 3: it consisted of a Kihara linear core potential and an electrostatic potential between point quadrupoles fixed at the CM of each molecule. The lattice constant $a$ at zero temperature, $T = 0$ K, was calculated to be 5.5372$_5$ Å,[1] while an experimental lattice constant is 5.5544 Å.[4,5] In Ref. 5, the lattice constants $a$ between 20 and 114 K are expressed by $a = 5.540 + 4.679 \times 10^{-6} T^2$ in units of Å by experiment. From these results, the lattice constants at finite temperatures were assumed to be $a = 5.5372_5 + 4.679 \times 10^{-6} T^2$ in the present work. The pressure was not calculated in the present work but assumed to be ambient as the temperature dependence of the lattice constant $a$ used in the MC simulations was based on the experiment at ambient pressure as described above. A theoretically consistent method is to undertake the pressure-constant MC simulations, but that is an issue to be studied in the future. Note also that solid $CO_2$ (dry ice) sublimates at 194.7 K at ambient pressure, while solid $CO_2$ at 200 K was simulated in the present work. This does not seem to alter the major results in the present work as the potential model used does not have such an accuracy as to quantitatively reproduce the sublimation temperature though it needs to be proved. In the simulations of the $CO_2$ crystal without vacancy, the CM of the $CO_2$ molecule at the origin was fixed all the time to prevent the possibility of a parallel drift of the entire crystal. Likewise, the vacancy was fixed at the origin all the time. These fixtures prevented the entire crystal from rotation for both cases with and without vacancy.

Table 1 shows the calculated energies per molecule (not divided by two) averaged over 40 data for different temperatures. The results for the $CO_2$ crystal without vacancy are listed in the column of "No vacancy", while those with a vacancy are listed in the column of "Single vacancy". The notations for the energies are explained at the bottom of the table. The energies are expressed in K units, where 1 K = 1.38054 × 10$^{-23}$ J. Note that the values in E(*Pa3*)NN in the column of "No vacancy" are identical to the value of E(*Pa3*) as E(*Pa3*)NN includes the interaction energy with the molecule at the origin. The MC simulation data in Table 1 are also depicted in Fig. 1 as a function of *a*. The uppermost curve is the result of the $CO_2$ crystal with a vacancy and the overlapped lower three curves are the average molecular energies of the 12 NN molecules of the $CO_2$ crystals without vacancy, and the average molecular energies for the $CO_2$ crystals with and without vacancy. The energy increase with temperature is the largest for the nearest neighbor (NN) molecules adjacent to the vacancy, implying that the molecules are more movable owing to the space due to the vacancy than those in the bulk that are more tightly confined in a space surrounded by 12 NN molecules.

In the following, the results on the NN molecules will be described. The NN molecules are labelled according to the molecular numbers used in the actual computation, and are shown in Fig. 2. Table 2 shows the positional and orientational deviations of the NN molecules measured from the perfect *Pa3* structure. The upper table shows the deviations in the $CO_2$ crystal without vacancy while the lower table shows the deviations in the $CO_2$ crystal with a vacancy. Here, $\Delta R$ is an average over 40 data of $[(\Delta x)^2 + (\Delta y)^2 + (\Delta z)^2]^{1/2}$ for the CM of each molecule (*x*, *y*, *z*). For $T = 0$, 100, and 200 K in the upper table, $\Delta R$ are less than 0.06 Å, 1.5% of the NN distance. On the other hand, $\Delta \Omega$ is an average angular deviation of the molecules from the *Pa3* structure. The values are less than or equal to 2.5°. On the other hand, in the lower table for the $CO_2$ crystal with a single defect, $\Delta R$ are less than 0.08 Å. A notable feature is seen in the magnitude of $\Delta \Omega$ in which the values for molecules 2, 3, and 4 as well as their equivalent molecules 255, 1824, and 2018 (see Fig. 2) are greater than other molecules at $T = 0$ and 100 K. This feature has already been seen in Ref. 1. More specifically, the molecules in close proximity to the oxygen atoms that belong to the molecule removed from the origin to make a vacancy change their orientations $\Delta \Omega$ by 3.4° ~ 4.2° at $T = 0$ and 100 K. These numbers are hatched in green in the lower table of Table 2. By contrast, $\Delta \Omega$ were very similar, between 1.7° ~ 2.0°, for all NN molecules at $T = 200$ K. This indicates that at least below $T = 100$ K, the molecular configurations around the vacancy are distorted similarly to those at $T = 0$ K. However, the differences in molecular orientations among the different NN molecules eventually disappeared at a temperature between $T = 100$ and 200 K. It was not determined whether such a transition takes place gradually with temperature or more suddenly in a narrow temperature range.

A comparison between the present results and those in Ref. 1 indicated that at $T = 0$ K, there was no significant difference in the molecular energies and the changes in the configurations of the NN molecules in the $CO_2$ crystals with and without vacancy. By contrast, the standard deviation in the molecular energy for the $CO_2$ crystal without vacancy was 4.1 K in Ref. 1, while it was 54.2 K in the present work. The major reason for the difference is most likely due to the difference in the values of $\{\Delta r, \Delta \psi\}$ used in the MC simulations. Preliminary examinations indicated that this was also the case for MC simulations at finite temperatures. The small values of $\{\Delta r, \Delta \psi\}$ in Ref. 1 were selected to determine an energetically stable structure, and the values of $\{\Delta r, \Delta \psi\}$ in the present work were determined to allow large molecular motions at finite temperatures: in both cases, the values of $\{\Delta r, \Delta \psi\}$ are "arbitrary". It therefore follows, unphysically, that thermodynamic properties such as specific heat depends on the "arbitrary" determined parameters $\{\Delta r, \Delta \psi\}$. Unfortunately, no solution to solve the problem has been found yet. So far, MC simulations have been successfully used for studying structural and thermodynamic properties of molecular fluids and biomolecules but there are only few studies on simple non-spherical molecular crystals[7] presumably because of the problem mentioned above. In the present work, it is concluded at this stage that MC simulation is useful for structural study of molecular solids, but there is an issue to be solved that energy fluctuation depends on the "arbitrary" parameters $\{\Delta r, \Delta \psi\}$. Finally, it should be mentioned that a crude estimate shows that the librational amplitude is ~2.7° and the positional amplitude is ~0.1 Å in quantum mechanics. Since the quantum effects are significant roughly below 1/10 of the Debye temperature,[6] *i.e.*, below 20 K for the present case of solid $CO_2$, special attentions are needed in interpreting the results of MC simulations at low temperatures.

In summary, molecular configurations were studied using Metropolis MC simulation, and it was found that the distorted configuration at low temperature around the vacancy became uniform between 100 and 200 K. It was also found that the standard deviation in the crystal energy strongly depended on the "artificial" parameters $\{\Delta r, \Delta \psi\}$ so that it is an important issue to determine the physically sensible values for $\{\Delta r, \Delta \psi\}$ in the study of simple molecular crystals by MC simulation.

Table 1. Calculated energies per molecule. The energies are in units of K.

| Temp. (K) | Lattice constant (Å) | Pa3 without vacancy | | | | Single defect | | | |
|---|---|---|---|---|---|---|---|---|---|
| | | E(Pa3) (K) | E(MC) (K) | E(Pa3)NN (K) | E(MC)NN (K) | E(Pa3) (K) | E(MC) (K) | E(Pa3)NN (K) | E(MC)NN (K) |
| 0 | 5.5573 | -6206.7 | -6171.5 | -6206.7 | -6172.5 | -6203.6 | -6180.2 | -5814.8 | -5808.9 |
| 100 | 5.5840 | -6190.3 | -6163.7 | -6190.3 | -6166.1 | -6187.2 | -6160.6 | -5793.5 | -5789.0 |
| 200 | 5.7243 | -5987.2 | -5960.9 | -5987.2 | -5961.0 | -5984.3 | -5958.0 | -5590.8 | -5555.3 |

E(*Pa3*): Static energy per molecule in the *Pa3* structure

E(MC): Average energy per molecule as a result of MC simulation

E(*Pa3*)NN: Static energy per molecule of the NN molecules in the *Pa3* structure

E(MC)NN: Average energy per molecule of the NN molecules as a result of MC simulation

Table 2. Positional and orientational deviations of the NN molecules measured from the *Pa3* structure. The upper graph shows the deviations in the crystal structure without vacancy while the lower graph shows the deviations in the crystal structure with a vacancy. In both cases, the deviations are measured from the perfect *Pa3* structure.

| Molecule label | *Pa3* orientation | No vacancy | | | | | |
|---|---|---|---|---|---|---|---|
| | | 0 K | | 100 K | | 200 K | |
| | | Δr | ΔΩ | Δr | ΔΩ | Δr | ΔΩ |
| 2 | $<\bar{1}11>$ | 0.0046 | 1.8 | 0.0101 | 1.8 | 0.0231 | 2.2 |
| 3 | $<1\bar{1}1>$ | 0.0058 | 1.8 | 0.0197 | 1.7 | 0.0389 | 1.9 |
| 4 | $<11\bar{1}>$ | 0.0099 | 1.8 | 0.0118 | 1.5 | 0.0311 | 2.1 |
| 31 | $<1\bar{1}1>$ | 0.0388 | 1.8 | 0.0062 | 1.9 | 0.0156 | 1.7 |
| 32 | $<11\bar{1}>$ | 0.0411 | 1.6 | 0.0162 | 1.6 | 0.0436 | 2.2 |
| 226 | $<\bar{1}11>$ | 0.0349 | 1.7 | 0.0085 | 1.8 | 0.0460 | 1.9 |
| 227 | $<1\bar{1}1>$ | 0.0257 | 1.8 | 0.0247 | 1.8 | 0.0303 | 2.0 |
| 255 | $<1\bar{1}1>$ | 0.0348 | 1.8 | 0.0172 | 1.7 | 0.0608 | 2.1 |
| 1794 | $<\bar{1}11>$ | 0.0476 | 2.0 | 0.0129 | 1.5 | 0.0356 | 1.8 |
| 1796 | $<11\bar{1}>$ | 0.0448 | 1.8 | 0.0274 | 1.6 | 0.0401 | 2.0 |
| 1824 | $<11\bar{1}>$ | 0.0557 | 1.7 | 0.0123 | 1.7 | 0.0280 | 1.9 |
| 2018 | $<\bar{1}11>$ | 0.0587 | 1.9 | 0.0094 | 1.4 | 0.0227 | 2.5 |
| Average | | 0.0335 | 1.8 | 0.0147 | 1.7 | 0.0347 | 2.0 |

| Molecule label | *Pa3* orientation | Single vacancy | | | | | |
|---|---|---|---|---|---|---|---|
| | | 0 K | | 100 K | | 200 K | |
| | | Δr | ΔΩ | Δr | ΔΩ | Δr | ΔΩ |
| 2 | $<\bar{1}11>$ | 0.0622 | 3.7 | 0.0576 | 3.4 | 0.0046 | 1.8 |
| 3 | $<1\bar{1}1>$ | 0.0701 | 3.8 | 0.0504 | 3.6 | 0.0058 | 1.8 |
| 4 | $<11\bar{1}>$ | 0.0565 | 3.7 | 0.0630 | 3.9 | 0.0099 | 1.8 |
| 31 | $<1\bar{1}1>$ | 0.0061 | 1.9 | 0.0378 | 2.3 | 0.0388 | 1.8 |
| 32 | $<11\bar{1}>$ | 0.0279 | 1.4 | 0.0451 | 1.9 | 0.0411 | 1.6 |
| 226 | $<\bar{1}11>$ | 0.0218 | 1.7 | 0.0360 | 2.3 | 0.0349 | 1.7 |
| 227 | $<1\bar{1}1>$ | 0.0371 | 1.8 | 0.0430 | 2.0 | 0.0257 | 1.8 |
| 255 | $<1\bar{1}1>$ | 0.0360 | 3.6 | 0.0771 | 3.8 | 0.0348 | 1.8 |
| 1794 | $<\bar{1}11>$ | 0.0496 | 1.9 | 0.0466 | 2.2 | 0.0476 | 2.0 |
| 1796 | $<11\bar{1}>$ | 0.0398 | 1.6 | 0.0126 | 1.8 | 0.0448 | 1.8 |
| 1824 | $<11\bar{1}>$ | 0.0385 | 3.6 | 0.0457 | 4.2 | 0.0557 | 1.7 |
| 2018 | $<\bar{1}11>$ | 0.0387 | 4.2 | 0.0440 | 3.6 | 0.0587 | 1.9 |
| Average | | 0.0404 | 2.7 | 0.0466 | 2.9 | 0.0335 | 1.8 |

*Figure captions*

Fig. 1. Calculated energies per molecule as a function of the lattice constant. The data points are the results of MC simulation, and the lines are only for eye guide. The energy is in units of K. For notations, see Table 1.

Fig. 2. Nearest-neighbor molecules around the vacancy at the origin.

Fig. 1. Calculated energies per molecule as a function of the lattice constant. The data points are the results of MC simulation, and the lines are only for eye guide. The energy is in units of K. For notations, see Table 1.

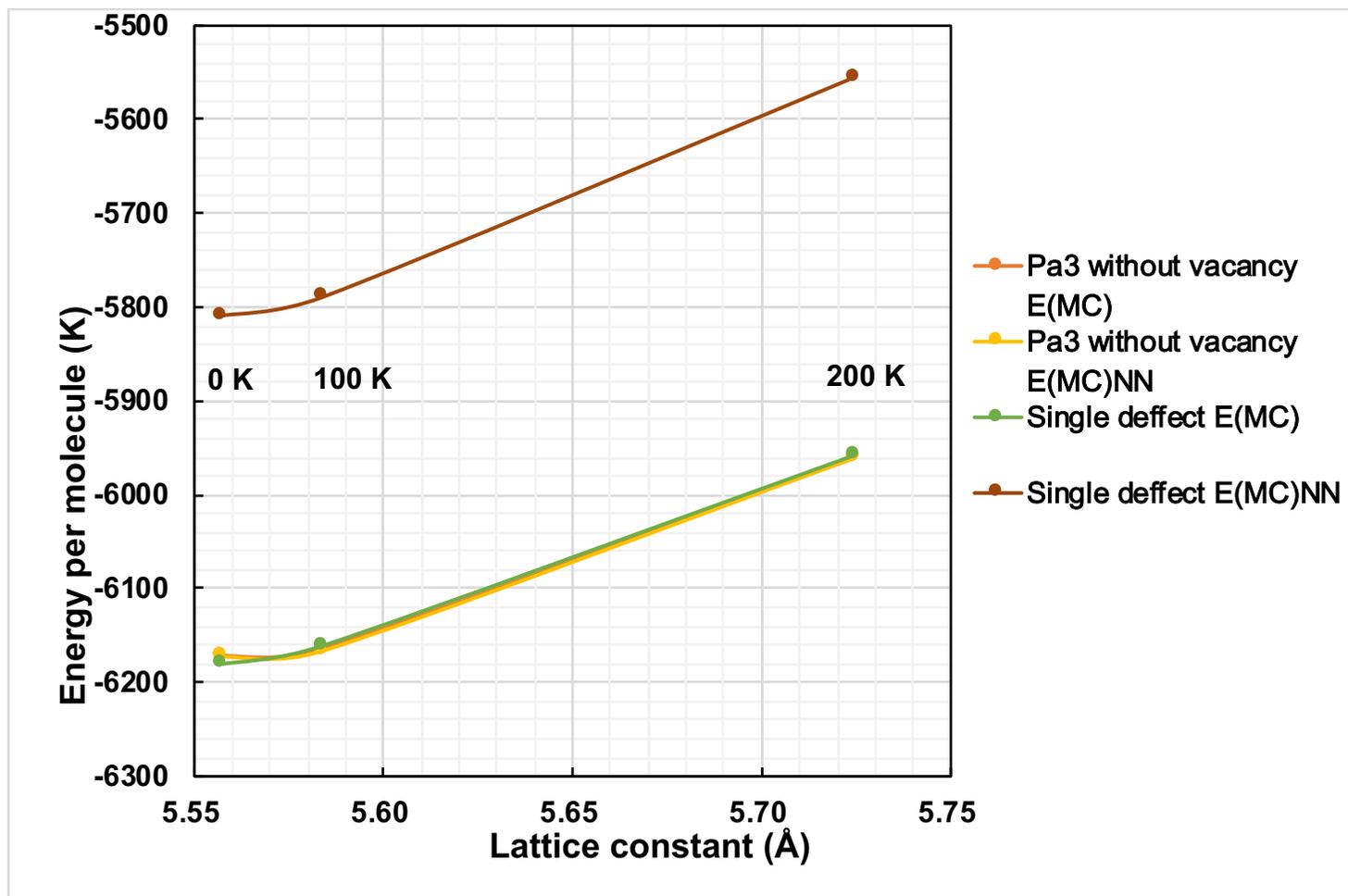

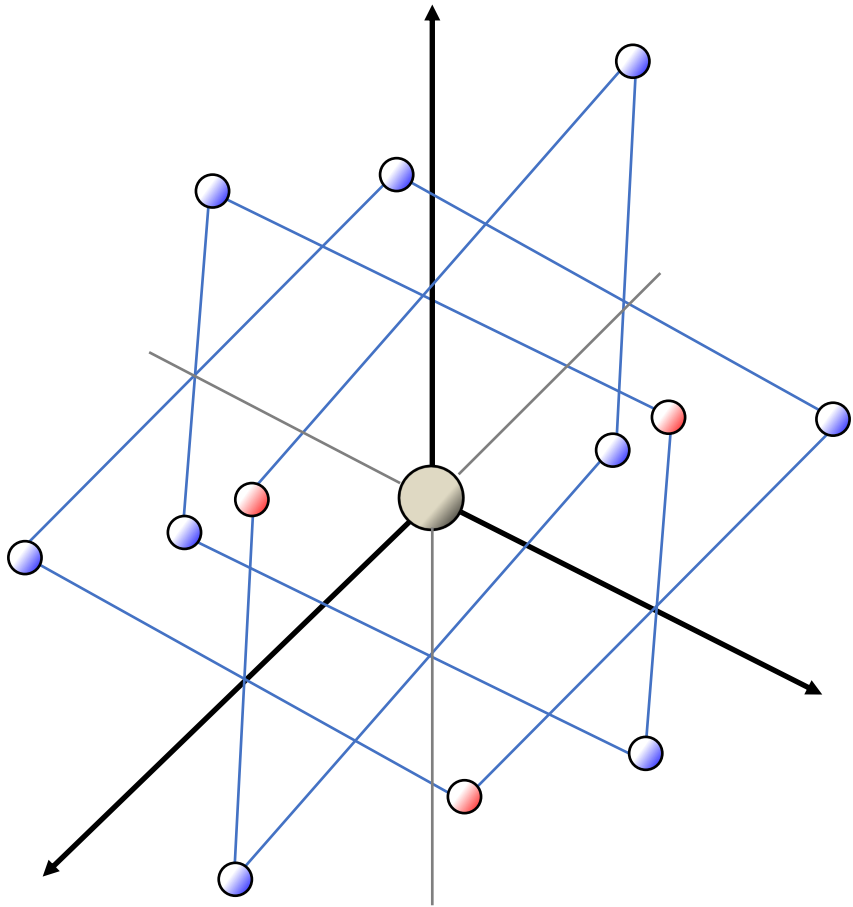

Fig. 2. Nearest-neighbor molecules around the vacancy at the origin.